\documentstyle[twocolumn,prb,aps,epsfig]{revtex}

\def\VEV#1{\left\langle #1\right\rangle}
\begin{document}
\title{
       Low-temperature transport in Heisenberg chains}

\author{J.V. Alvarez and Claudius Gros 
       } 
\address{Fakult\"at 7, Theoretische Physik,
 University of the Saarland,
66041 Saarbr\"ucken, Germany.}

\maketitle

\begin{abstract}
A technique to determine accurately transport
properties of integrable and non-integrable quantum-spin
chains at finite temperatures by Quantum Monte-Carlo
is presented. The reduction of the Drude weight
by interactions in the integrable gapless regime 
is evaluated. Evidence
for the absence of a Drude weight
in the gapless regime of a non-integrable system 
with longer-ranged interactions is presented. We estimate
the effect of the non-integrability on the
transport properties and compare with recent experiments
on one-dimensional quantum-spin chains.
     
\end{abstract}


\vspace*{1cm}
{\bf Introduction} -
During the last few years 
several families of materials containing
well characterized quasi one-dimensional 
spin-1/2 structures have been synthesized. 
The charge-transfer gap is in many 
cases large and the spin excitations 
contribute significantly to
the thermal and magnetization transport at low 
temperatures. For example, 
$^{63}$Cu NMR studies \cite{Tak96} 
in  Sr$_2$CuO$_3$ have
measured a spin diffusion coefficient (equivalent 
to diffusive magnetization transport) several  
orders of magnitude larger than the value for 
conventional diffusive systems, 
and thermal transport measurements in
Sr$_2$CuO$_3$ and SrCuO$_2$ indicate \cite{Sol00} 
quasi-ballistic transport with a mean-free
path of several thousands of $\AA$. 

These unusual results have been related to the peculiar
physics of one-dimensional quantum chains.
It is known that the spin transport in the 
XXZ chain
\[
H^{(xxz)}\ =\ {\sum}_{i} \left[\frac{J_{xx}}{2}
\left(S_{i}^{+}S_{i+1}^{-}+S_{i}^{-}S_{i+1}^{+}\right)
+ J_{z} S_{i}^{z}S_{i+1}^{z}\right]
\]
is not diffusive, even for $T\gg J_{xx}$ 
\cite{Nar96,Fab97}. A connection between integrability 
and transport 
\cite{Cas95,Zotos-Prelovsek96,Zotos-Naef-Prelovsek97,Naef98,Nar98} 
enlights the situation. In a generic integrable 
model, like $H^{(xxz)}$, the spin current is not conserved,    
but it has a non-vanishing `component' with respect
to the projection onto a conserved operator.
As a consequence, the current-current correlation 
functions do not decay to zero for large times.

The hamiltonian appropriate for real compounds 
like Sr$_2$CuO$_3$ and SrCuO$_2$ correspond
to $H^{(xxz)}$ only in first approximation.
A crucial question is therefore:
How do small deviations from 
integrability change the picture described above? 
Extending an earlier analysis by Giamarchi \cite{Gia92},
Rosch and Andrei concluded recently within a
memory-matrix approach \cite{ROSCH},
that deviations from
integrability lead to an exponentially large
conductivity in Hubbard-like models away from
commensurability.


Despite the on-going effort devoted to this problem,
the fundamental difference in between integrable and
non-integrable models have not yet shown up in
QMC-simulations \cite{SCALAPINO,HANKE}, presumable due to
the demanding numerical requirements.
In this Letter we develop the techniques and tools for
the data-analysis necessary for extracting 
the Drude weight for integrable and the lifetime, mean-free path and 
the diffusion constant for non-integrable quantum-spin models
from correlation functions obtained by
QMC simulations using the Loop Algorithm \cite{EVERTZ}.  
We present extensive comparison with exact results obtained
by Bethe-Ansatz and compare for a non-integrable system with
recent experiments on Sr$_2$CuO$_3$.

 
{\bf Drude weight} -
QMC-simulations yield in general correlations functions
on the imaginary-time axis. We therefore consider
the Kubo formula for the dynamical conductivity
$\sigma(\omega_n)=\lim_{q\to0}\sigma(q,\omega_n)$
\begin{equation}
\sigma(q,\omega_n)\ =\ 
{-\VEV{K} \,-\, \Lambda(q,\omega_{n})\over \omega_n}
\ \equiv\ {D(q,\omega_n)\over \omega_n}
\label{sigma}
\end{equation}
on the imaginary axis, 
where $\VEV{K}$, for the models we will consider,
 is the expectation value of the
kinetic energy per site and where
$\Lambda$ is the current-current correlation as a 
function of the Matsubara frequency,
\begin{equation}
  \Lambda(q,\omega_{n})\,=\, 
  \frac{1}{L} \int_{0}^{\beta} e^{i \omega_{n} \tau}
 \VEV{j^{z}(q,\tau)j^{z}(-q,0)}\,d\tau~.
 \label{Lambda}
\end{equation}
Eq.\ (\ref{sigma}) leads via
$\sigma(\omega)= \pi D(T)\delta(\omega)
+\sigma_{reg}(\omega)$ to the conventional relation 
\cite{SCALAPINO} 
between the current-current correlation function  
and the Drude weight (we are borrowing the terminology of 
electrical responses)
$D(T) =\lim_{\omega_{n\rightarrow 0}}\lim_{q\rightarrow0}D(q,\omega_{n})$ 
\begin{equation}
D(T) \ = \
 -\VEV{K} \,-\, \Lambda(q \rightarrow 0,\omega_{n}\rightarrow 0 )~.
\label{Drude}
\end{equation}
Note that $\Lambda(q,\omega+i\delta)$ is analytic in the
upper half of the complex $\omega$-plane and the extrapolation
along the imaginary axis can be reliably performed at low
temperatures, when many Matsubara-frequencies
$\omega_n=2\pi T n$ are available close to $\omega=0$ for the
extrapolation \cite{HANKE}.     
   
The continuity equation 
$
{\partial\over \partial t} S_l^z(t)\,+\, 
{\partial\over \partial x} j_l^z(t)  = 0
$
leads to the expression
\begin{equation}
 j_{l}^{z}\ =\ -iJ_{xx}(S_{l}^{+}S_{l+1}^{-}-S_{l}^{-}S_{l+1}^{+})
\label{j_z}
\end{equation}  
for the current operator $j_l(t)$ in the XXZ-model.
In Fourier space, the continuity equation 
takes the form
\begin{equation}
{d\over d\tau}S_q^z(\tau) \ =\
[H,S_q^z] \ =\ i\left( 1-e^{iq}\right) j_q^z~.
\label{cont_q}
\end{equation}  
$\Lambda(q,\omega_n)$ is a non-diagonal four-site operator.
In principle non-diagonal operators can be computed using 
the loop algorithm \cite{WIESE} but, as it was discussed 
recently \cite{ALVAREZ}, the algorithm to compute
two-site correlation functions is more 
efficient than the algorithms 
that compute operators involving three or more sites.   
To obtain high quality data we measure directly
the dynamical susceptibility in imaginary time
\begin{equation}
S(q,\omega_{n})\ =\ {1\over L}\int_{0}^{\beta}
e^{i \omega_{n} \tau}\langle S_{q}^{z}(\tau)S_{-q}^{z}(0) \rangle~,
\label{S_q_o}
\end{equation}
a simple diagonal two-site correlation function. 
The Drude peak is computed by using Eq.\ (\ref{Drude})
and the relation
\begin{equation}
%
\omega_n^2 \,S(q,\omega_{n})\ =\ 
4\sin^2(q/2)\, D(q,\omega_n)~.
\label{relation}
\end{equation}
This relation can be derived by a two-fold partial
integration of the right-hand side of Eq.\ (\ref{S_q_o})
with respect to $\tau$, which leads to
\[
S(q,\omega_{n})\ =\ {-1\over\omega_n^2}
\langle\, [[H,S_q^z],S_{-q}^z]\, \rangle
\,-\,{4\sin^2(q/2)\over\omega_n^2} \Lambda(q,\omega_n)~,
\]
where we have used Eq.\ (\ref{cont_q}) for the 
imaginary-time derivatives of $S_q^z$ and $S_{-q}^z$
and definition (\ref{Lambda}).
Evaluation of the double commutator $[[H,S_q^z],S_{-q}^z]$,
which is the boundary term from the 
partial integration,
then leads to Eq.\ (\ref{relation}). Note, that this
double commutator occurs here for the Matsubara
correlation functions and does not occur for a
related real-frequency correlation function \cite{Naef98}.

{\bf Data-Analysis} -
At low temperatures and frequencies, the scaling of $D(q,\omega_{n})$
can be obtained simply invoking the conformal symmetry 
of the model emerging  in the gapless regime $J_{z}<J_{xx}$.
$S(q,\omega_{n})$ at small q takes then form
$
 S(q,\omega_{n}) = D_1(T)q^{2}/((cq)^2+\omega_{n}^{2}).
$
This expression and Eq.\ (\ref{relation}) suggest the form
\begin{equation}  
 D(q,\omega_{n})\,=\,
 \frac{D_1(T)\,\omega_{n}^{2}}{\Delta^{2}(q)+\omega_{n}^{2}}~.
\label{LL}
\end{equation}
Alternatively, Eq.~(\ref{LL}) can be viewed as the first term
of the exact representation for $D(q,\omega_{n})$ containing
an infinite-number of terms \cite{HANKE}.

The XXZ-model maps to an interacting  1D spinless fermionic 
system at half filling. For the noninteracting case 
(the XX chain) we can compute exactly $D(q,\omega_{n})$
and we obtain $D(0)=J_{xx}/\pi$, 
and  $\Delta(q)=J_{xx}\sin(q)\sim J_{xx} q$. 
For $J_{z}<J_{xx}$ the Umklapp term
remains marginally irrelevant and one expects Luttinger
Liquid like  correlation functions \cite{Gia92} like
Eq.\ (\ref{LL}).  

In the ideal Luttinger Liquid the bosonic excitations 
are arbitrarily well defined at low temperatures  
and low frequencies.
The conductivity
is necessarily infinite in that case, and 
ansatz (\ref{LL}), being invariant under time reversal
reflects  that property.  Nevertheless, 
we are also interested in the study of more
general situations in which the bosons   
can decay, become quasiparticles,  
and memory effects can be taken into account.  
To this end we use the  formalism of the memory matrices  
which has been successfully applied to study how
one-dimensional electron liquids 
can gain a resistivity\cite{Gia92,ROSCH}.

This is achieved by taking 
the following fitting function for $D(q,\omega_n>0)$:
\begin{equation}     
D(q,\omega_{n})\ =\ \sum_{j=1}^{2}
\frac{D_j(q)\,\omega_{n}^{2}}
{\Delta_j^{2}(q)+2\gamma_j(q)\,\omega_n+ \omega_{n}^{2}}~.
 \label{fitting}
\end{equation}    
$D(q,\omega_{n})$ is analytic in the upper complex-plane
for $\gamma_j(q)\ge0$. 
For $\Delta_i(0)=\lim_{q\to0}\Delta_i(q)$ we find
either a

(i) gapless regime:
$\ \Delta_1(0)=0\ $ and $\ \Delta_2(0)>0$,  or a

(i) gap-full regime:
$\ \Delta_1(0)>0\ $ and $\ \Delta_2(0)>\Delta_1(0)$.\\
The first term in Eq.\ (\ref{fitting}) dominates the
low-frequency behavior in both cases and we have set
generally $\gamma_2\equiv0$ in order to keep the
number of parameters to a minimum.

In the gapless regime the optical conductivity (\ref{sigma})
takes for small frequencies the Drude form
\begin{equation}
\mbox{Re}\,\sigma(\omega)\ =\ 
{2D_1(0)\gamma_1(0)\over \omega^2+4\gamma_1^2(0)}
\ \equiv\
{\sigma_0\over 1 + (\omega\tau)^2}~,
\label{Drude_formula}
\end{equation}
where we introduced the
DC-conductivity $\sigma_0 = D_1(0)/(2\gamma_1(0))$
and the quasi-particle lifetime
$\tau = (2\gamma_1(0))^{-1}$.
For $\tau\to\infty$
Eq.\ (\ref{Drude_formula}) reduces to
$\mbox{Re}\,\sigma(\omega)= \pi D_1(0)\,\delta(\omega)$.
At high frequencies 
\begin{equation}     
\lim_{\omega_n\to\infty}D(0,\omega_{n})\ =\  -\VEV{K}\ \equiv\
D_1(0)+ D_2(0)
\label{limit_large_omega_n}
\end{equation}
and a finite $D_2(0)$ results in a reduction of 
the Drude weight $D(T)$ with respect to the kinetic energy,
see Eq.\ (\ref{Drude}).
A finite $D_2(0)$ measures therefore the amount of
decay experienced by the {\em total} current due to the interactions.  
We note that the Ansatz
Eq.\ (\ref{fitting}) for $D(q,\omega_n)$, 
together with Eq.\ (\ref{limit_large_omega_n}), 
is consistent with the f-sum-rule \cite{Bae79}
$\int_0^\infty \mbox{Re}\,\sigma(\omega) = -{\pi\over 2}\VEV{K}$
for the optical conductivity.

\begin{figure}[t]
\noindent
\\
\centerline{
\epsfig{file=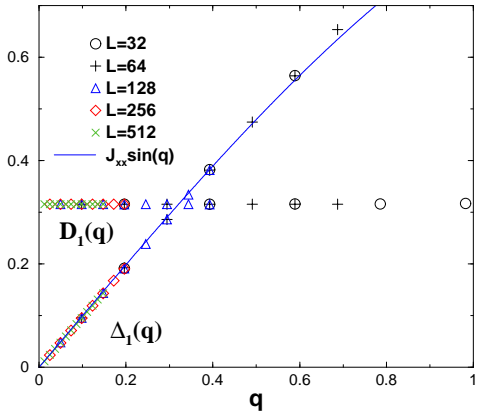,width=0.35\textwidth} 
}
\caption{\label{fig1}
The parameters $D_1(q)$ and $\Delta_1(q)$ from Eq.\ (\ref{fitting})
as a function of momenta $q$,
for the XX-model ($J_{z}=0$) and various system sizes
at $T=0.004\,J_{xx}$. Statistical error bars are of the order of the 
system size \protect\cite{note_1}.
}
           
\end{figure}

{\bf Ballistic transport} -
In Fig.\ \ref{fig1}  we show the values 
for $D_1(q)$ and $\Delta_1(q)$
for the XX chain for different system sizes as obtained
by QMC\cite{note_1}, $D_2(q)$ and $\gamma_1(q)$ optimize to zero for
this model.  The different system sizes collapse at the smallest 
momenta and in this way the thermodynamic limit and the
$q \rightarrow 0$ limit are performed simultaneously.
Our prescription to extract $D_1(0)$ as the limiting
value $D_1(0)=\lim_{q\to0}D_1(q)$ is more involved
than previous $q\equiv0$ calculations \cite{HANKE}.
but we believe it to be necessary in order to avoid substantial finite-size
effects due to the fact that $D(q\to0,\omega_n)$ becomes a step-function
at $\omega_n=0$ in the thermodynamic limit, compare Eq.~(\ref{LL}).

\begin{figure}[t]
\centerline{
\epsfig{file=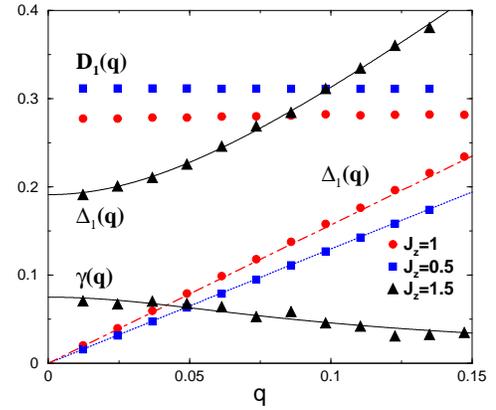,width=0.35\textwidth} 
           }
\vspace{4pt}
\caption{\label{fig2}
$D_1(q)$, $\Delta_1(q)$ and
$\gamma_1(q)$ from Eq.\ (\ref{fitting})
as a function of momenta $q$
for the XXZ-model, $L=512$ and 
for various $J_{z}$ at $T=0.004\,J_{xx}$.
$\gamma_1(q)$ is too small for $J_z\le J_{xx}$ to
show up on this scale.
The lines are the Bethe Ansatz result (\ref{c_exact})
for the velocity $c(J_z)$ (no fit, for $J_z\le J_{xx}$).
For the discussion of the fit for $J_z=1.5 J_{xx}$ see
the text.
}
\end{figure}

For finite interaction $J_z$ we find
in the gapless regime $J_z\le J_{xx}$, that
$\Delta_1(q)\approx c(J_z)\,q$
where $c(J_{z})$ is the velocity 
\begin{equation}
c(J_z) \ =\ {\pi\over2}\,
{\sqrt{J_{xx}^2-J_z^2}\over 
\mbox{arcos}(J_{z}/J_{xx})}
\label{c_exact}
\end{equation}
of the des Cloiseaux-Pearson spectrum \cite{Cloi66}.
In Fig.\ \ref{fig2} we compare the measured value 
of $c(J_z)$ with the Bethe Ansatz result (\ref{c_exact})
for $J_{z}/J_{xy}=1/2,1$.
$\Delta_1(q)$ is fitted well in the gapped phase
by $\varepsilon(q)=\sqrt{\Delta_0^2+(cq)^{2}}$. We find
$\Delta_0=0.191\,J_{xx}$, which is close to twice the 
one-magnon gap of $0.091\,J_{xx}$ \cite{Cloi66}.

The damping $\gamma_1(q)$ is vanishing small
for $J_{z}<J_{xx}$ and acquires a finite value in
the gapped phase which can be fitted
phenomenological by the relation 
$\ \gamma_1(q)\Delta_1(q)\approx\,const.$, 
independent of $q$. 

\begin{figure}[t]
\centerline{
\epsfig{file=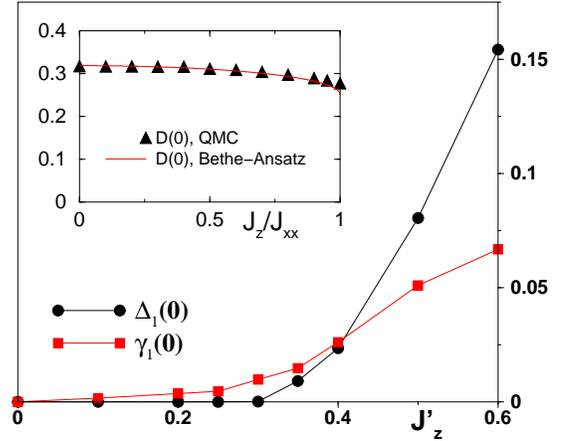,width=0.40\textwidth} 
           }
\vspace{4pt}
\caption{\label{fig3}
For $L=512$ and $T=0.004\,J_{xx}$ the
QMC-results for the gap $\Delta_1(0)$ and the relaxation
rate $\gamma_1(0)$ as a function of $J_{z}' $
for $H^{(xxz)}+H'$ with $J_z=J_{xx}\cos(\pi/6)$. 
Inset: The QMC-results for the Drude weight for 
$J_{z}'=0 $ and $T=0.004\,J_{xx}$ and $L=512$
in comparison with the Bethe-Ansatz
result at $T=0$, Eq.\ (\ref{D_exact}). Note that $D(T)$
is smooth for low T (see Fig.\ \protect\ref{fig4}). 
}
\end{figure}

In Fig.\ \ref{fig3} (inset) we present values obtained by QMC
for the Drude weight in the gapless regime,
$J_z<J_{xx}$ at $T=0.004\,J_{xx}$. We find good
agreement with the $T=0$ Bethe Ansatz result \cite{SHASTRY}
\begin{equation}
D(0) \ =\ {\nu^2\over 4\pi} {\sin(\pi/\nu)\over \nu-1}~,\qquad
J_z=J_{xx}\cos(\pi/\nu)~,
\label{D_exact}
\end{equation}
specially at the smaller values of $J_z$.
In this region the correlation function is expected to 
satisfy more accurately the scaling law (\ref{LL}).  
The $S(q,\omega_{n})$ in the isotropic Heisenberg model
does not follow strictly the Luttinger Liquid form.
It presents multiplicative logarithmic corrections,  
that should be observed also for values of $J_z$ slightly smaller 
than $J_{xx}$  when  finite temperatures and system sizes impose
a cut-off for the RG equation of the Umklapp 
coupling constant \cite{Gia92}.
   
\begin{figure}[t]
\centerline{
\epsfig{file=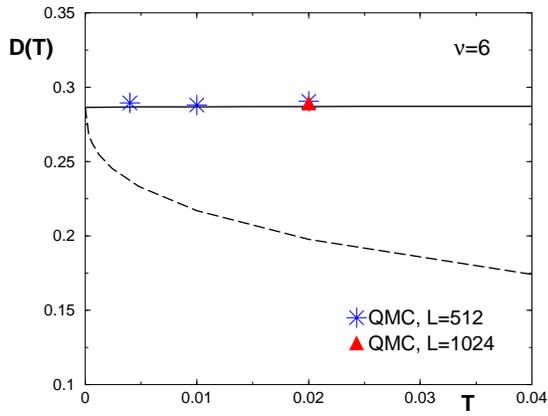,width=0.40\textwidth} 
           }
\vspace{4pt}
\caption{\label{fig4}
QMC results for the Drude weight for 
$L=512,1024$ and $J_z=J_{xx}\cos(\pi/6)$
as a function of temperature (in units of $J_{xx}$)
in comparison with two (solid lines: Ref.\ \cite{KLUEMPER},
dashed lines: Ref.\ \cite{ZOTOS99}) Bethe-Ansatz results.
}
\end{figure}

We study now the behavior of the Drude weight at finite
temperatures for models free from strong 
multiplicative corrections. 
The main conclusion of Zotos \cite{ZOTOS99} 
is the fast decay 
of the Drude weight when the temperature increases,
%
$
D(T)-D(0)\ \sim\  -\kappa\, T^{\frac{2}{\nu-1}}
$
%
where $\kappa$ is a constant and $\nu$ defined by
Eq.\ (\ref{D_exact}). This rapid decrease with
increasing temperature is consistent with
exact diagonalization studies at hight temperatures \cite{Nar98}.
Kl\"umper {\it et al.} have found on the other hand \cite{KLUEMPER}, 
with an alternative Bethe-ansatz approach, a functionally different
behavior for $D(T)$, see Fig.\ \ref{fig4}.
For a numerical probe of $D(T)$ we focus
on $\nu=6$ and consider several small temperatures.  
In Fig.\ (\ref{fig4}) we show a comparison of our data with
the two available analytical results \cite{ZOTOS99,KLUEMPER}.
Our results agree with the temperature-dependence predicted
by Kl\"umper {\it et al.}, suggesting in particular a finite
Drude weight also at the isotropic point \cite{KLUEMPER}.

{\bf Diffusive transport} -
We consider now a perturbation of $H^{(xxz)}$   
that breaks the integrability but keeps the $z$-component
of the magnetization conserved:
\begin{equation}
H'\ =\ J'_{z}\,{\sum}_{i}S^{z}_{i}S^{z}_{i+3}~.
\label{Hnonint}
\end{equation}
The expression (\ref{j_z}) for the spin current remains
valid.
For this model, $H=H^{(xxz)}+H'$,
we predict a transition to a gapped phase around
$J_z'\cong 0.3\,J_{xx}$ (for $\nu=6$), see Fig.\ \ref{fig3}. 
We find the relaxation time 
$\tau=1/2\gamma_1(0)=\lim_{q\to0}1/2\gamma_1(q)$ to be finite 
within the numerical accuracy (due to finite-$q$ and $\omega_n$
resolution), leading to a finite DC-conductivity in the
gapless phase.

For $1/\omega\gg \tau$ the optical 
conductivity takes (for small $cq/\gamma_1(0)$) the 
diffusion form
\begin{equation}
\sigma(q,\omega)\ =\ 
{\sigma_{0}\,\omega\over \omega +i D_s\,q^2}~,\quad
D_s = {c^2\over2\gamma_1(0)}\equiv c^2\tau~.
\label{diffusion}
\end{equation}
$D_s$ is the spin-diffusion constant. Eq.\ (\ref{diffusion})
is consistent with $D_s=c\lambda_s$, where
$\lambda_s=c\tau$ is the mean free length. 

We have evaluated $\sigma(T)$,
in addition to the data presented in Fig.\ \ref{fig3},
for $\nu=6,\ J_z' = 0.3 \,J_{xx}$ and
find
 $\sigma(T=0.004J_{xx})=13.6 \pm 0.9$, 
 $\sigma(T=0.008J_{xx})=12.1 \pm 1.0$ and
 $\sigma(T=0.012J_{xx})=10.1 \pm 0.8$.

We now take $J_{xx}=2000\,\mbox{K}$, which is appropriate for
Sr$_2$CuO$_3$ and evaluate the transport coefficients
for $J_z'=0.3\,J_{xx},\ \nu=6,\ T=0.004\,J_{xx}$.
We find $\lambda \approx 88$ lattice constants
and $D_s\approx 6\times10^{15}\,\mbox{sec}^{-1}$.
We do not expect this model to be directly relevant for
Sr$_2$CuO$_3$. But if we ask ourselves now the
question, whether the experimental results
for $\lambda$ and $D_s$ being large (but finite)
could be explained within a pure spin-model with a
small deviation from integrability, we might expect
$H=H^{(xxz)}+H'$ to show the characteristic behavior
of a non-integrable quantum-spin chain. If we now
change the deviation from integrability 
(controlled by $J_z'$)
such that $\gamma_1(0)$ decreases by a factor of about
ten, then both $\lambda$ and $D_s$ would increase
by the same factor and would be consistent 
with the experimentally measured values \cite{Tak96,Sol00}.
 

{\bf Discussion} -
We have shown that certain transport properties of
quantum-spin chains can be evaluated directly from
two-point correlations functions using a relation
in between $S(q,\omega_n)$ and the the dyamical 
conductivity which we have derived.
For the integrable chains we
support the original suggestion by Zotos {\it et al.}
\cite{Cas95,Zotos-Prelovsek96,Zotos-Naef-Prelovsek97}
of a finite Drude weight at finite temperatures and
settle a recent dispute regarding the functional form
of $D(T)$. In addition we presents results suggesting
the absence of ballistic transport (i.e.\ zero Drude-weight)
for a non-integrable model, for which we are able
to estimate magnitude of the DC-conductivity.
We have shown that the experiments
on quasi one-dimensional spin-compounds \cite{Tak96,Sol00} 
are, in principle, consistent with the notion that they
probe directly non-integrability effects, 
though we cannot rule out at this point
that disorder or the coupling to phonons \cite{Nar96} would
lead to the observed finite values for the transport coefficients.
  

{\bf Acknowledgments} -
This work was supported by the DFG. We  
would like to thank A. Kl\"umper and X. Zotos for giving us their
numerical results and their valuable comments.We also 
want to thank A.Sologubenko and R. Valent{\'\i}
a careful reading and several suggestions.




\end{document}